# Origin of the metal-to-insulator crossover in cuprate superconductors

F. Laliberté[1,2], W. Tabis[2,3], S. Badoux[1], B. Vignolle[2], D. Destraz[4], N. Momono[5], T. Kurosawa[6], K. Yamada[7], H. Takagi[8], N. Doiron-Leyraud[1], C. Proust[2,9] and Louis Taillefer[1,9]

*1 Département de physique & RQMP, Université de Sherbrooke, Sherbrooke, Québec J1K 2R1, Canada*

*2 Laboratoire National des Champs Magnétiques Intenses (CNRS, INSA, UJF, UPS), Toulouse 31400, France*

*3 AGH University of Science and Technology, Faculty of Physics and Applied Computer Science, 30-059 Krakow, Poland*

*4 Department of Physics, University of Zürich, 8057 Zürich, Switzerland*

*5 Department of Applied Sciences, Muroran Institute of Technology, Muroran 050-8585, Japan*

*6 Department of Physics, Hokkaido University, Sapporo 060-0810, Japan*

*7 Institute of Materials Structure Science, High Energy Accelerator Research Organization & The Graduate University for Advanced Studies, Oho, Tsukuba 305-0801, Japan*

*8 Department of Advanced Materials, University of Tokyo, Kashiwa 277-8561, Japan*

*9 Canadian Institute for Advanced Research, Toronto, Ontario M5G 1Z8, Canada*

**Superconductivity in cuprates peaks in the doping regime between a metal at high**
**$p$ and an insulator at low $p$. Understanding how the material evolves from metal to**
**insulator is a fundamental and open question[1,2]. Early studies in high magnetic**
**fields revealed that below some critical doping an insulator-like upturn appears in**
**the resistivity of cuprates at low temperature[3,4], but its origin has remained a**
**puzzle[5]. Here we propose that this "metal-to-insulator crossover" is due to a drop**
**in carrier density $n$ associated with the onset of the pseudogap phase at a critical**
**doping $p*$. We use high-field resistivity measurements on $La_{2-x}Sr_xCuO_4$ to show**
**that the upturns are quantitatively consistent with a drop from $n = 1 + p$ above $p*$**
**to $n = p$ below $p*$, in agreement with high-field Hall data in $YBa_2Cu_3O_y$ (ref. 6).**
**We demonstrate how previously reported upturns in the resistivity of**
**$La_{2-x}Sr_xCuO_4$ (ref. 3), $YBa_2Cu_3O_y$ (ref. 5) and $La_{1.6-x}Nd_{0.4}Sr_xCuO_4$ (ref. 7) are**
**explained by the same universal mechanism: a drop in carrier density by 1.0 hole**
**per Cu atom.**

At high doping, the Fermi surface of cuprates is a large hole-like cylinder with
a carrier density $n = 1 + p$ (ref. 8). Consequently, the normal-state Hall coefficient $R_H$
measured at $T \to 0$, in magnetic fields large enough to suppress superconductivity,
yields a Hall number $n_H = V / e \, R_H = 1 + p$ (refs. 7,9,10), where $e$ is the electron charge
and $V$ the volume per Cu atom in the $CuO_2$ planes. Above a critical doping $p*$ (Fig. 1),
the normal-state electrical resistivity $\rho$ decreases monotonically upon cooling towards
$T = 0$, with $p* = 0.18$ in $La_{2-x}Sr_xCuO_4$ (LSCO) (refs. 3,11), $p* = 0.19$ in $YBa_2Cu_3O_y$
(YBCO) (refs. 6,12), and $p* = 0.235$ in $La_{1.6-x}Nd_{0.4}Sr_xCuO_4$ (Nd-LSCO) (refs. 7,13). In
LSCO at $p = 0.23$ (ref. 11) and Nd-LSCO at $p = 0.24$ (ref. 7), $\rho(T)$ is seen to decrease
linearly as $T \to 0$. By contrast, below $p*$, $\rho(T)$ develops an upturn as $T \to 0$ (refs. 3, 7).
This change from linear-$T$ decrease to upturn has been called a "metal-to-insulator
crossover" (refs. 3, 4). Its origin has remained a puzzle until now, with tentative
explanations invoking localization[3,4], the Kondo effect[5], or stripe order[14], for example.

In Nd-LSCO at $p = 0.20$, the upward deviation in $\rho(T)$ from its linear $T$ dependence at high $T$ starts at a temperature $T^* = 80 \pm 10$ K (ref. 7). Photoemission (ARPES) measurements[15] show that the pseudogap opens at this same temperature $T^*$. The same study finds that there is no pseudogap at $p = 0.24$, where there is indeed no resistivity upturn[7]. Similarly, ARPES in LSCO yields $T^* = 130 \pm 20$ K at $p = 0.15$ (ref. 16), in agreement with the start of the upturn in $\rho(T)$ (ref. 17), and no pseudogap is detected at $p = 0.22$ (ref. 16), in agreement with the lack of upturn[11]. These studies establish the link between resistivity upturn and pseudogap.

In this paper, we focus on the onset of the pseudogap at $T = 0$, upon crossing $p^*$, and are not concerned with the nature and the impact of the inelastic scattering in the high-temperature pseudogap phase or the actual value of $T^*$. A key signature of the pseudogap phase at $T = 0$ was recently revealed by high-field measurements of $R_H$ in YBCO: $n_H$ drops from $1 + p$ to $p$ upon crossing below $p^*$ (ref. 6). Using this insight, we show that the metal-to-insulator crossover in cuprates can be explained quantitatively in terms of a drop in carrier density from $n = 1 + p$ to $n = p$, caused by a $T = 0$ metal-to-metal transition into the pseudogap phase at $p^*$.

The in-plane resistivity of LSCO was measured in four samples, with dopings $p = 0.136$, $0.143$, $0.157$ and $0.163$, respectively. The zero-field curves of $\rho(T)$ are displayed in Fig. 2 (and Fig. S1). Isotherms of $\rho$ vs $H$ obtained from field sweeps up to 56 T (or 58 T), at various temperatures, are displayed in Fig. S2. In Fig. 3, we plot the six lowest isotherms at $p = 0.136$; we see that the normal-state resistivity increases systematically as temperature is reduced from 45 K down to 1.5 K. In Fig. 2, we plot $\rho$ at $H = 55$ T vs $T$. These normal-state data (at 55 T) reveal a pronounced upturn below $\sim 50$ K, consistent with prior data at similar dopings[3,4]. As $T \to 0$, $\rho(T)$ saturates to a finite value $\rho(0)$.

The size of the upturn decreases with doping, and it disappears at $p^* = 0.18$

(Fig. 1), where $\rho(T)$ decreases linearly as $T \to 0$ (ref. 11). To quantify the size of the upturn, we introduce $\rho_0$ , the residual resistivity the sample would have if there were no upturn (no pseudogap), obtained by fitting the data at high $T$ to $\rho(T) = \rho_0 + a\, T$ (Fig. 2 and Fig. S1). Indeed, high-field data in LSCO at $p = 0.21$ (ref. 11), just above $p^*$, shows that a linear fit to $\rho(T)$ between 150 K and 200 K does yield the correct value of the residual resistivity, a measure of the disorder in the sample, when extrapolated to $T = 0$. Given that data as a reference, we apply the same fitting procedure to our data at $p < p^*$. Reasonably, we find that $\rho_0 \sim 40\ \mu\Omega$ cm for all 4 samples (Fig. S1).

We define the quantity $n_\rho = (1 + p)\, \rho_0 / \rho(0)$, which by construction is the carrier density above $p^*$. Indeed, $n_\rho = 1 + p$ at $p = 0.18$, 0.21 and 0.23, three dopings where no upturn is observed in LSCO (ref. 11). In Fig. 4a, we plot $n_\rho$ vs $p$. We observe that $n_\rho$ drops precipitously below $p^*$, to reach $n_\rho \sim p$ at $p \sim 0.15$, within an interval $\delta p \sim 0.03$ (Fig. 4a). In Fig. 4b, we compare the drop of $n_\rho$ in LSCO to the drop of $n_H$ in YBCO. The similarity is striking, with $n_H$ also dropping from $1 + p$ to $p$ within an interval $\delta p \sim 0.03$ below $p^*$. We propose that the drops in $n_H$ and $n_\rho$ are both caused by a drop in carrier density from $n = 1 + p$ to $n = p$ at $p^*$. In other words, the mechanism for the upturn in the resistivity of cuprates is a loss of carrier density caused by the onset of the pseudogap phase. We can then account quantitatively for the previous data on LSCO, Nd-LSCO and YBCO.

In Fig. 4a, we add two points obtained from early data on LSCO, at $x = 0.15$ and $x = 0.17$ (ref. 3) (Fig. S3). They fit in very well with our own, such that all points fall on a smooth line connecting $n_\rho = 1 + p$ at $p = 0.18$ to $n_\rho = p$ at $p = 0.143$. In Fig. 4b, we also plot $n_\rho$ obtained from data in Nd-LSCO (ref. 7), and observe a transition from $n_\rho = 1 + p$ at $p = 0.24$ to $n_\rho = p$ at $p = 0.20$, of width $\delta p \sim 0.03$. In addition, we plot the Hall number in Nd-LSCO (ref. 7), and see that $n_H = n_\rho$ within error bars (Fig. 4b). This shows that the mobility $\mu \sim R_H / \rho$ does not change appreciably through the transition, so that $\rho = 1 / (n \, e \, \mu)$ and $R_H = V / (n \, e)$ both scale simply as $1 / n$. In the Supplementary Information, we use the magneto-resistance in our data to show that the mobility in LSCO also does not change appreciably across $p^*$. This confirms that it is indeed a drop in carrier density $n$ that drives the upturns in $\rho(T)$ and in $R_H(T)$.

For YBCO, it has not been possible so far to measure the normal-state $\rho(T)$ as $T \to 0$ because the fields needed to fully suppress superconductivity exceed 115 T in the relevant doping range[18], *i.e.* between $p = 0.16$ and $p = 0.19$. Using electron irradiation to disorder a YBCO sample with oxygen content $y = 7$ ($p = 0.18$), and hence lower its critical field, Rullier-Albenque and co-workers[5] were able to measure $\rho(T)$ as $T \to 0$. They saw an upturn with $\rho(0) / \rho_0 = 2.0 \pm 0.1$ (ref. 5), so that $n_\rho = 0.59 \pm 0.03$, in good agreement with the $n_H$ data on YBCO (Fig. 4b). Note that in YBCO, as in Bi-2201, it has not yet been established that $\rho(T)$ is linear at $p^*$ as $T \to 0$, so it is not clear that we can apply the procedure used for LSCO and Nd-LSCO to extract the value of $\rho_0$. Note also that unlike in LSCO and Nd-LSCO, $\rho(T)$ in YBCO and Bi-2201 drops below $T^*$, at least initially[17], further complicating the quantitative analysis of any low-$T$ upturn. Nevertheless, the values of $\rho(0)$ measured in Bi-2201 are consistent with a drop of carrier density from $n = 1 + p$ to $n = p$ (Fig. S7).

We conclude that the transition into the pseudogap phase at $T = 0$ upon crossing below $p^*$ has a universal signature: the loss of 1.0 hole per Cu in the carrier density $n$, which goes from $n = 1 + p$ to $n = p$. This is a metal-to-metal transition that does not involve any insulator or localization. Indeed, at the end of the transition, where $n \sim p$, the resistivity does not diverge as $T \to 0$, whether in LSCO at $p = 0.136$ (Fig. S1a) or YBCO at $p = 0.18$ (ref. 5) or Nd-LSCO at $p = 0.20$ (ref. 7). The reported $\log T$ behavior is only observed at lower doping[3,4], when $\rho(0)$ becomes large enough that $k_F \, l \sim 1$ (Supplementary Information). The transition also does not involve the Kondo effect, as the upturns are seen in both $R_H(T)$ and $\rho(T)$. Finally, the transition at $p^*$ has nothing to do with charge-density-wave (CDW) order. Indeed, CDW order ends at a critical doping distinctly below $p^*$ (Fig. 1), as established for YBCO (ref. 6) and LSCO (refs. 19,20).

The abruptness of the drop in $n$ vs $p$ at $p^*$ (Fig. 4) points to a sharp transition as a function of doping at $T = 0$, rather than a crossover. This is confirmed by comparing to a theoretical calculation of $R_H$ for a hole-doped cuprate undergoing a $2^{nd}$ order quantum phase transition at $T = 0$ into a phase of long-range antiferromagnetic order[21], across $p_{AF} = 0.2$. In Fig. 4b, we see how the calculated Hall number agrees perfectly, within error bars, with the YBCO data. In the calculation, the Fermi surface below $p = 0.17$ consists of small nodal hole pockets whose volume is such that $n = p$, as imposed by the Luttinger rule. (That $n_H$ is slightly larger than $n = p$ is due to the fact that the hole pockets are not isotropic[21].) Between $p = 0.17$ and $p_{AF} = 0.2$, there is an intermediate regime, of width $\delta p = 0.03$, where the Fermi surface includes both nodal hole pockets and small electron pockets at antinodal locations[21].

It has been argued that the pseudogap phase can transform the Fermi surface into small hole pockets with $n = p$ without breaking translational symmetry (refs. 22,23,24). In the model of Yang, Rice and Zhang[22], the strong Umklapp scattering of the Mott insulator causes a $T = 0$ transition at $p^*$, where the large Fermi surface is transformed into small nodal hole pockets with $n = p$. As in the antiferromagnetic scenario above, this transformation also proceeds via an intermediate regime containing antinodal electron pockets, and calculations also agree well with the Hall data in YBCO (ref. 21).

## Methods

**Samples.** Large single crystals of LSCO were grown by the flux-zone technique, with nominal Sr concentrations of $x = 0.144$ and $0.15$ at Tohoku University, $x = 0.145$ at Hokkaido University, and $x = 0.16$ at the University of Tokyo. The long rods that are produced typically have a variation of Sr concentration along their length, so that the hole concentration (doping) $p$ at a particular point may not be equal to the nominal (average) value of $x$. Samples for resistivity measurements were cut in the shape of small rectangular platelets, of typical dimensions 1 mm × 2 mm × 0.5 mm, with the smallest dimension along the $c$ axis. Contacts were made using H20E silver epoxy, diffused by annealing. The superconducting transition temperature $T_c$ of the four samples was determined as the temperature below which the zero-field resistance is zero. The values are: $T_c$ = 36.0 K ($x$ = 0.144), 37.3 K ($x$ = 0.145), 36.8 K ($x$ = 0.15), and 36.1 K ($x$ = 0.16).

**Hole concentration.** The hole concentration (doping) $p$ was determined for each sample using the doping dependence of the (tetragonal to orthorhombic) structural transition temperature, $T_{LTO}$. The signature of $T_{LTO}$ in the resistivity is a small but sharp kink[17]. The values for our four samples are: $T_{LTO}$ = 214 K ($x$ = 0.144), 197 K ($x$ = 0.145), 163 K ($x$ = 0.15), and 147 K ($x$ = 0.16). We use the linear doping dependence of the anomaly reported in ref. 17, in the range $0.10 < x < 0.17$, to convert $x$ into $p$. We obtain $p$ = 0.136 ($x$ = 0.144), 0.143 ($x$ = 0.145), 0.157 ($x$ = 0.15), and 0.163 ($x$ = 0.16).

**Resistivity measurements.** The longitudinal resistance $R$ was measured in Sherbrooke in steady fields up to 16 T and in Toulouse in pulsed fields up to 58 T, with the field oriented along the $c$ axis. The pulsed-field measurements were performed with a current excitation between 5 mA and 10 mA at a frequency in the range 20-60 kHz. A high-speed acquisition system was used to digitize the reference signal (current) and the voltage drop across the sample at a frequency of 500 kHz. The data were post-analyzed with software to perform the phase comparison. Data for the rise and fall of the pulse were in good agreement, thus excluding any heating due to eddy currents. Tests at different frequencies showed excellent reproducibility.

**Error bars.** The uncertainty on the value of $p$ comes from the uncertainty in determining $T_{LTO}$ from the kink in $\rho(T)$, which we estimate to be ± 5 K. This translates into an uncertainty on $p$ of ± 0.002. Because we only use the ratio of $\rho(0)$ over $\rho_0$, the uncertainty on their absolute value, which comes from measuring the geometric factor of the samples, cancels out, $e.g.$ in the formula for calculating $n_\rho$. The error bar on $n_\rho$ comes from the uncertainty associated with extrapolating $\rho(T)$ to get $\rho(0)$, which we estimate to be ± 5 %, and with fitting $\rho(T)$ between 150 K and 200 K to get $\rho_0$, which we estimate to be approximately ± 20 %. The total error bar on $n_\rho$ is therefore approximately ± 25 %. The values of $\rho_0$ we obtain for our samples are all very close, ranging from 32 to 43 $\mu\Omega$ cm, consistent with the actual residual resistivity of LSCO samples with $p > p^*$, namely in the range from 10 to 50 $\mu\Omega$ cm (refs. 3,11).

**Magneto-resistance.** In Fig. 4a, we plot $n_\rho$ obtained by taking the raw value of $\rho(0)$ measured at $H = 55$ T, not only for our four samples (red squares), but also for the two samples in ref. 3 (red diamonds). In principle, one should correct these values for the magneto-resistance (MR) in the data. This is done in the Supplementary Information, where we show that correcting for the MR at $p = 0.136$ and $p = 0.143$ (Figs. S4 and S5) yields values of $n_\rho$ that are only slightly different (Fig. S6) and, as such, does not affect any of our conclusions. For $p = 0.157$ and $p = 0.163$, it is not possible to correct for the MR as higher fields would be needed to fully reach the normal state at the lowest temperatures (Fig. S2). Nevertheless, we expect the magnitude of the MR to be comparable to the MR at $p = 0.136$ and $p = 0.143$ since all samples have a comparable $\rho_0$, and hence a comparable level of disorder scattering, and so a comparable mobility at $T \rightarrow 0$.

**Sample size.** No statistical methods were used to predetermined sample size.

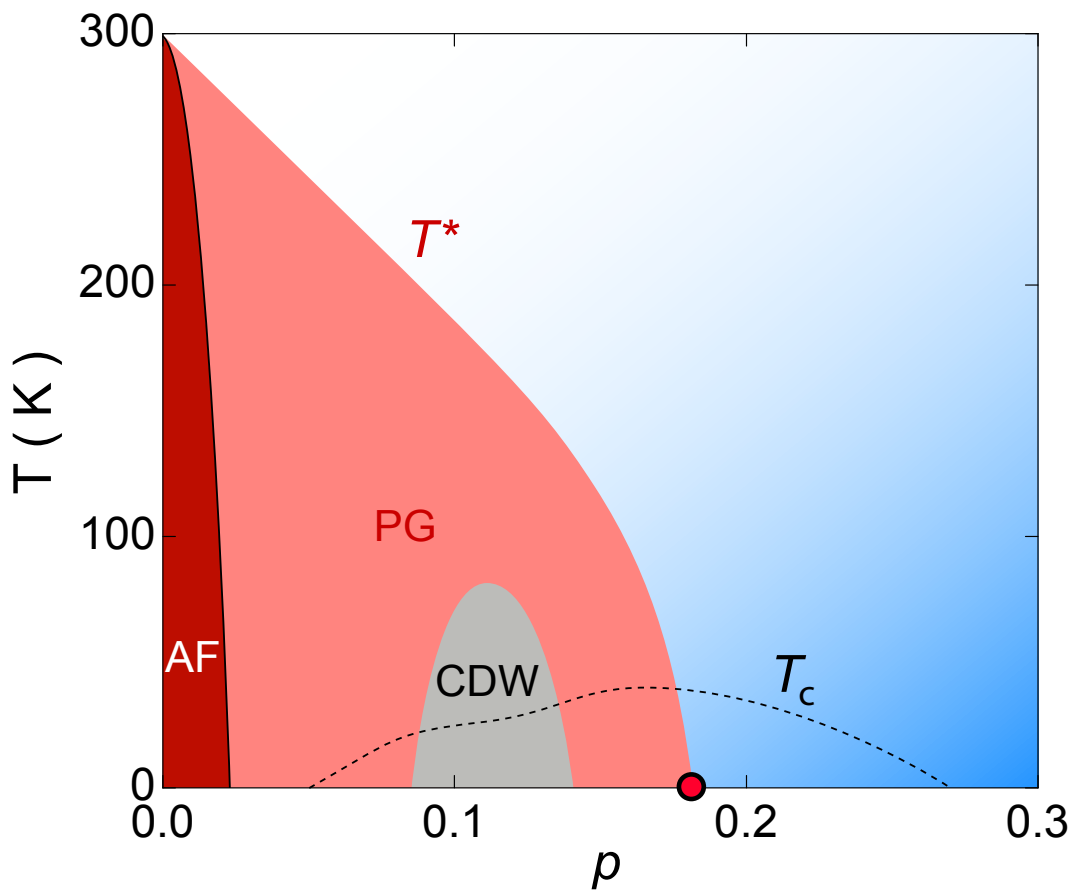

**Fig. 1 | Temperature-doping phase diagram of cuprates.**

Schematic phase diagram of hole-doped cuprates, consisting of four phases: antiferromagnetism (AF) at low doping (dark red); the pseudogap phase (light red), bounded by the crossover temperature $T^*$; charge-density-wave (CDW) modulations in the dome-like region centered at $p \sim 0.12$ (grey); superconductivity, delineated by the zero-field critical temperature $T_c$ (dashed line). By applying large magnetic fields, superconductivity is removed, revealing the critical point $p^*$ at which the pseudogap phase ends in the normal state at $T = 0$ (red dot). The CDW phase ends at a $T = 0$ critical doping distinctly below $p^*$ (refs. 6,20). Above $p^*$, the Fermi surface in the metallic phase at low temperature (blue) is a large hole-like cylinder with a carrier density $n = 1 + p$. In LSCO, $p^* = 0.18$ (refs. 3,11).

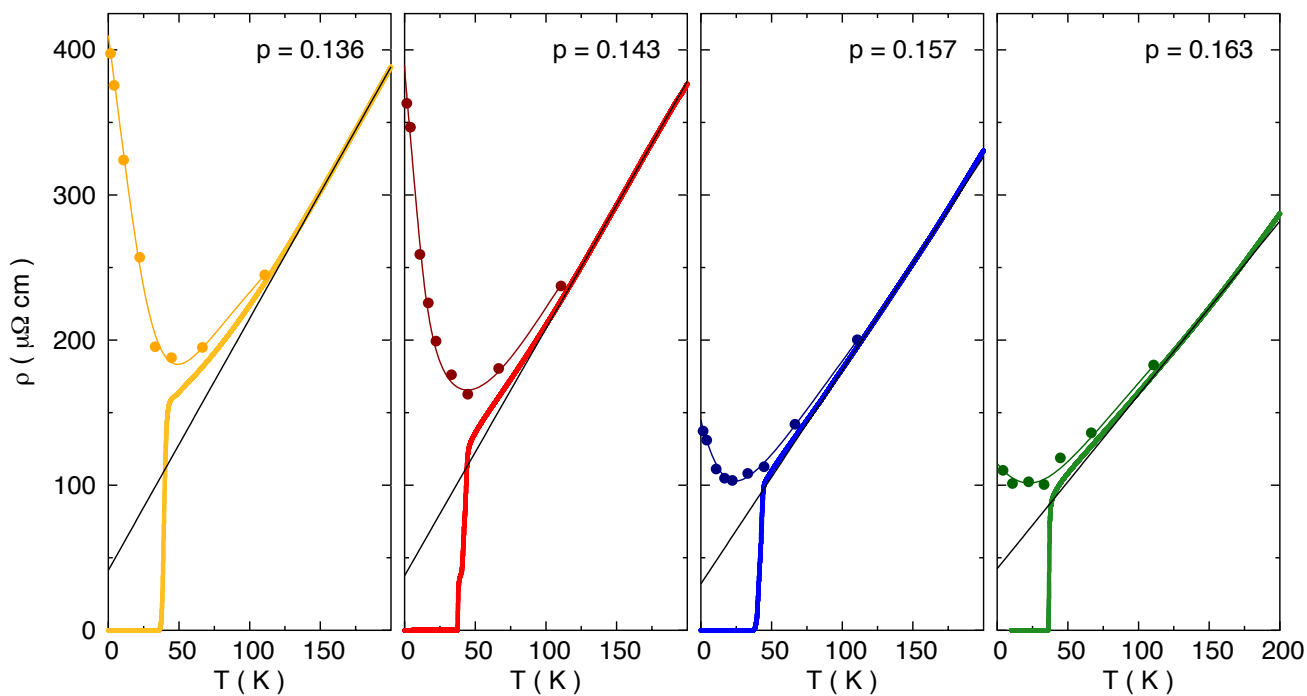

**Fig. 2 | Temperature dependence of the resistivity in LSCO.**

In-plane electrical resistivity ρ of our four LSCO samples as a function of temperature, with dopings *p* as indicated. The continuous curve is data taken in zero magnetic field (*H* = 0). The dots are the value of ρ at *H* = 55 T, obtained from isotherms of ρ vs *H* (Fig. S2). The thin line through the dots is a guide to the eye, whose value at *T* = 0 gives ρ(0) (Fig. S1). The straight black line is a linear fit to the zero-field data at high temperature, whose extrapolation to *T* = 0 gives ρ₀ (Fig. S1). The data deviate from the linear fit below the pseudogap temperature *T*\* (Fig. 1). This deviation grows to gradually develop into a large upturn as *T* → 0, the signature of what has been called the "metal-to-insulator crossover" of cuprates (refs. 3,4). The size of the upturn, measured by the ratio ρ(0) / ρ₀ (see text), is seen to decrease with doping.

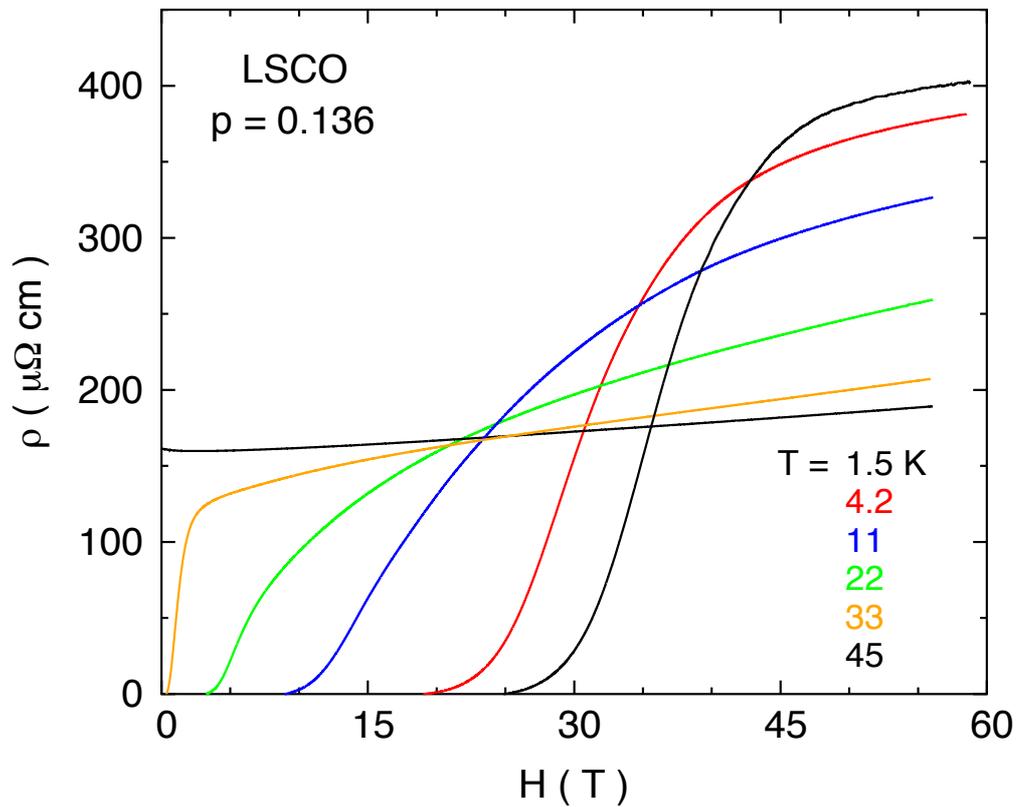

**Fig. 3 | Field dependence of the resistivity in LSCO.**

Isotherms of the resistivity ρ as a function of magnetic field *H* in our LSCO sample with doping *p* = 0.136 at different temperatures, as indicated. The normal-state resistivity (at high field) is seen to grow with decreasing temperature. The magneto-resistance at high field is discussed and analyzed in the Supplementary Information (Figs. S4 to S6). The isotherms for all four samples are displayed in Fig. S2. The value at *H* = 55 T is plotted vs temperature in Fig. 2, for all samples.

**a**

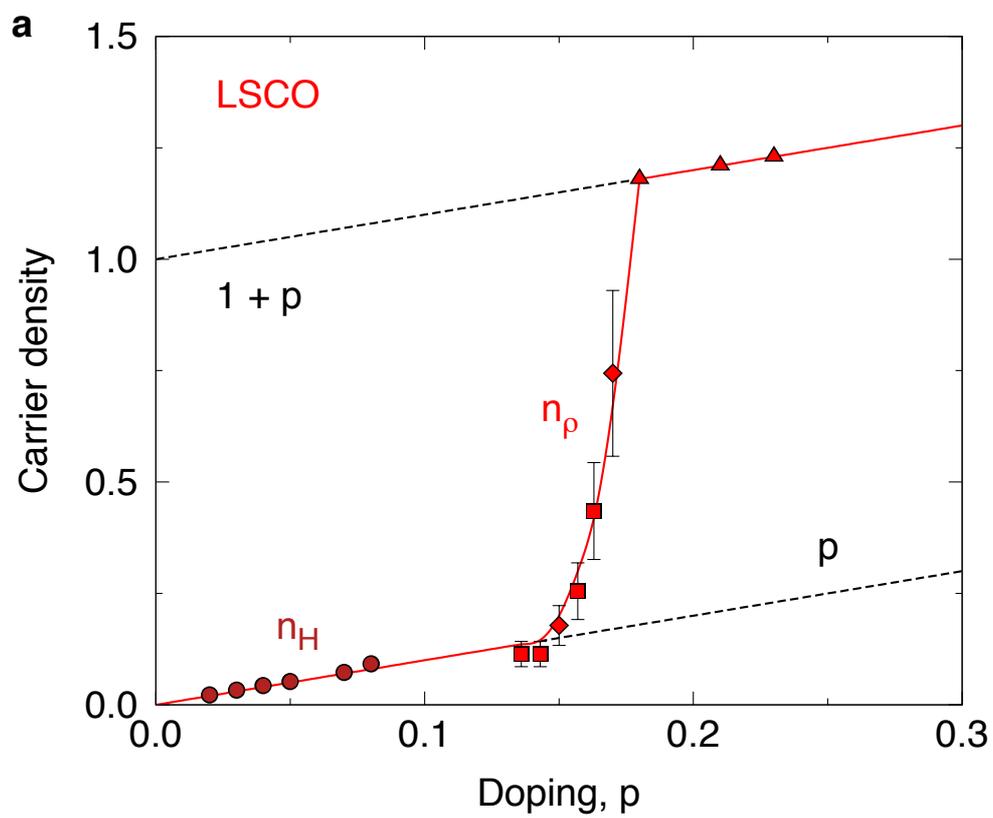

**b**

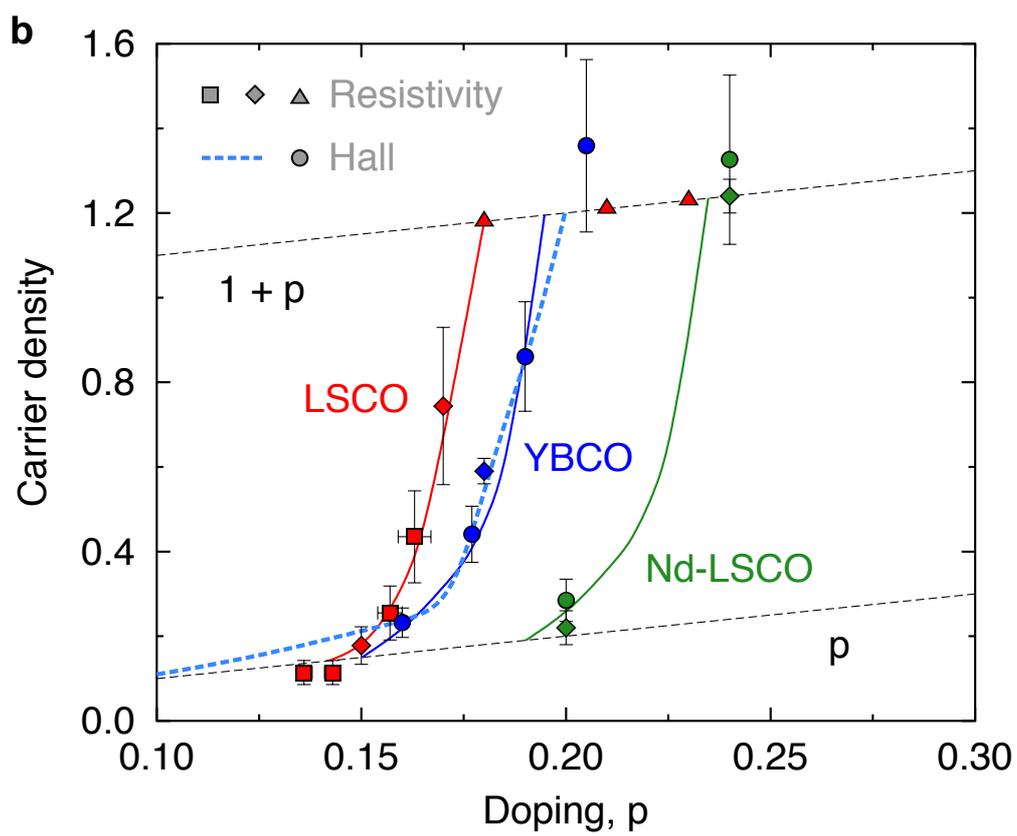

## Fig. 4 | Doping evolution of the normal-state carrier density.

Doping dependence of the normal-state carrier density $n$ of cuprates at low temperature, evaluated in two different ways: 1) from the relation $n_\rho = (1 + p)\, \rho_0 / \rho(0)$ (see text), using high-field measurements of the resistivity $\rho$ (squares, diamonds, and triangles); 2) from the Hall number $n_H = V / e\, R_H$ (circles). **a)** $n_\rho$ in LSCO (red squares, this work; red diamonds[3], Fig. S3; red triangles[11]); $n_H$ in LSCO, measured at $T = 50$ K (circles[25]). The solid red line is a guide to the eye. The dashed lines mark $n = 1 + p$ (upper) and $n = p$ (lower). The error bars reflect the uncertainty in extrapolating $\rho(T)$ to obtain $\rho_0$ (see Methods). **b)** Data on $n_\rho$ are shown for LSCO (red symbols, from panel a)), YBCO (blue diamond[5]) and Nd-LSCO (green diamonds[7]). Data on $n_H$ are from high-field measurements of $R_H$ in YBCO (blue circles[6]) and Nd-LSCO (green circles[7]). The solid red, blue and green lines are guides to the eye. With decreasing $p$, the carrier density is seen to drop rapidly from $1 + p$ to $p$ at $p^*$, in all three materials, with $p^* = 0.18$ (LSCO), $0.19$ (YBCO) and $0.235$ (Nd-LSCO). The dashed blue line is a calculation[21] of $n_H$ at $T = 0$ for a hole-doped cuprate undergoing a transition at $p = 0.2$ into a phase of antiferromagnetic order, with wavevector $\boldsymbol{Q} = (\pi, \pi)$.

# SUPPLEMENTARY INFORMATION

**Origin of the metal-to-insulator crossover in cuprate superconductors**

F. Laliberté, W. Tabis, S. Badoux, B. Vignolle, D. Destraz, N. Momono, T. Kurosawa, K. Yamada, H. Takagi, N. Doiron-Leyraud, C. Proust and Louis Taillefer

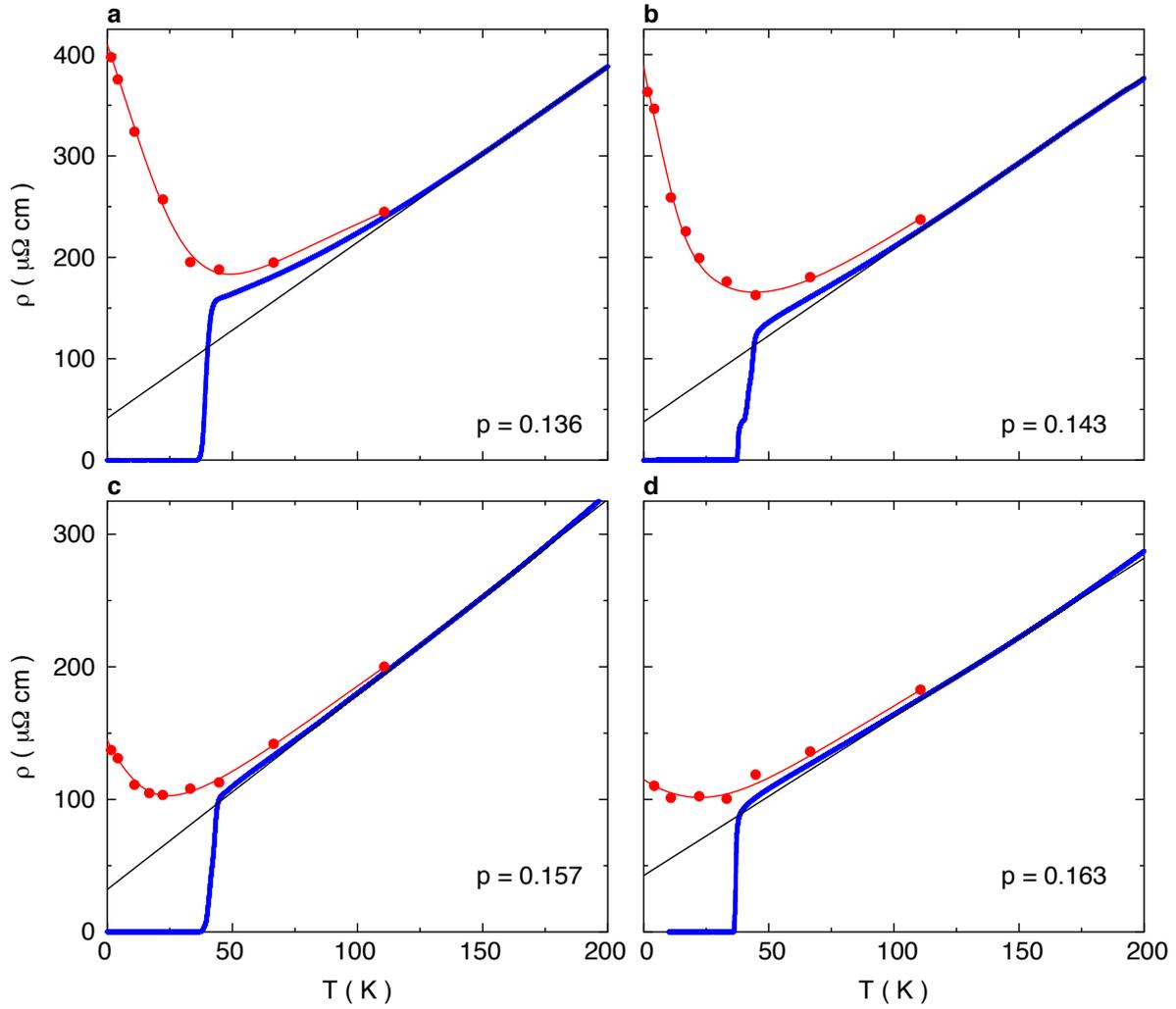

**Fig. S1 | Temperature dependence of the resistivity in LSCO.**
In-plane electrical resistivity ρ of LSCO as a function of temperature, for our four samples, with dopings as indicated: **a)** $p$ = 0.136 ($x$ = 0.144); **b)** $p$ = 0.143 ($x$ = 0.145); **c)** $p$ = 0.157 ($x$ = 0.15); **d)** $p$ = 0.163 ($x$ = 0.16). The continuous blue curve is data taken in zero magnetic field ($H$ = 0). The red dots are the values of ρ at $H$ = 55 T, obtained from isotherms of ρ vs $H$ (Fig. S2). The red line is a guide to the eye, whose value at $T$ = 0 is ρ(0). The straight black line is a linear fit to the zero-field data at high temperature. Its extrapolation to $T$ = 0 gives $\rho_0$. The values of ρ(0) and $\rho_0$ are used to calculate $n_\rho$ (see text), plotted as red squares in Fig. 4. They are : a) ρ(0) = 410 ± 20 μΩ cm, $\rho_0$ = 42 ± 8 μΩ cm; b) ρ(0) = 385 ± 20 μΩ cm, $\rho_0$ = 40 ± 8 μΩ cm; c) ρ(0) = 140 ± 10 μΩcm, $\rho_0$ = 32 ± 6 μΩ cm; and d) ρ(0) = 115 ± 10 μΩ cm, $\rho_0$ = 43 ± 9 μΩ cm.

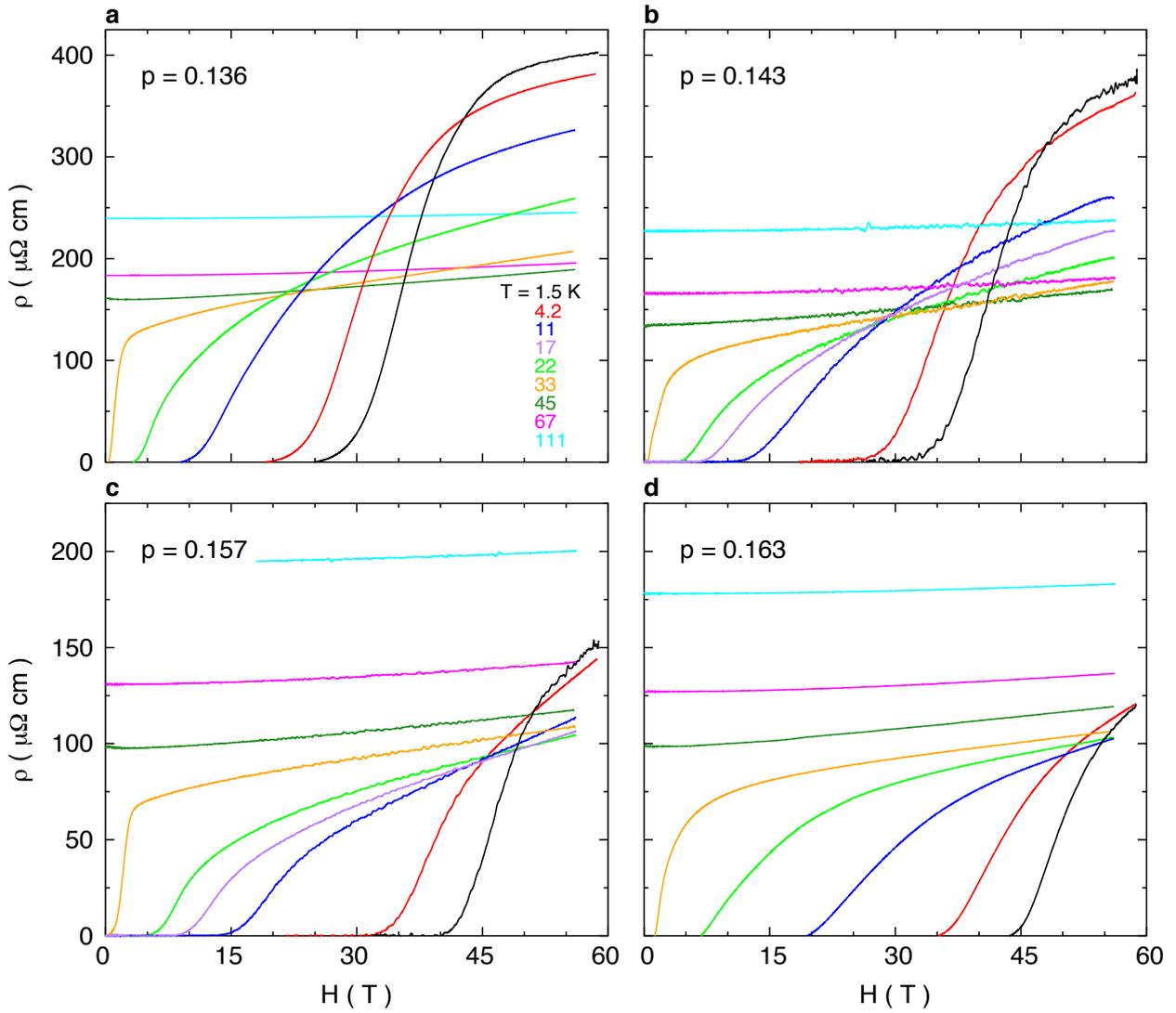

**Fig. S2 | Field dependence of the resistivity in LSCO.**

Isotherms of the resistivity ρ as a function of magnetic field $H$ at different temperatures, as indicated, in our four LSCO samples: **a)** $p$ = 0.136 ($x$ = 0.144); **b)** $p$ = 0.143 ($x$ = 0.145); **c)** $p$ = 0.157 ($x$ = 0.15); **d)** $p$ = 0.163 ($x$ = 0.16). The value at $H$ = 55 T is plotted vs temperature in Fig. 2 and Fig. S1.

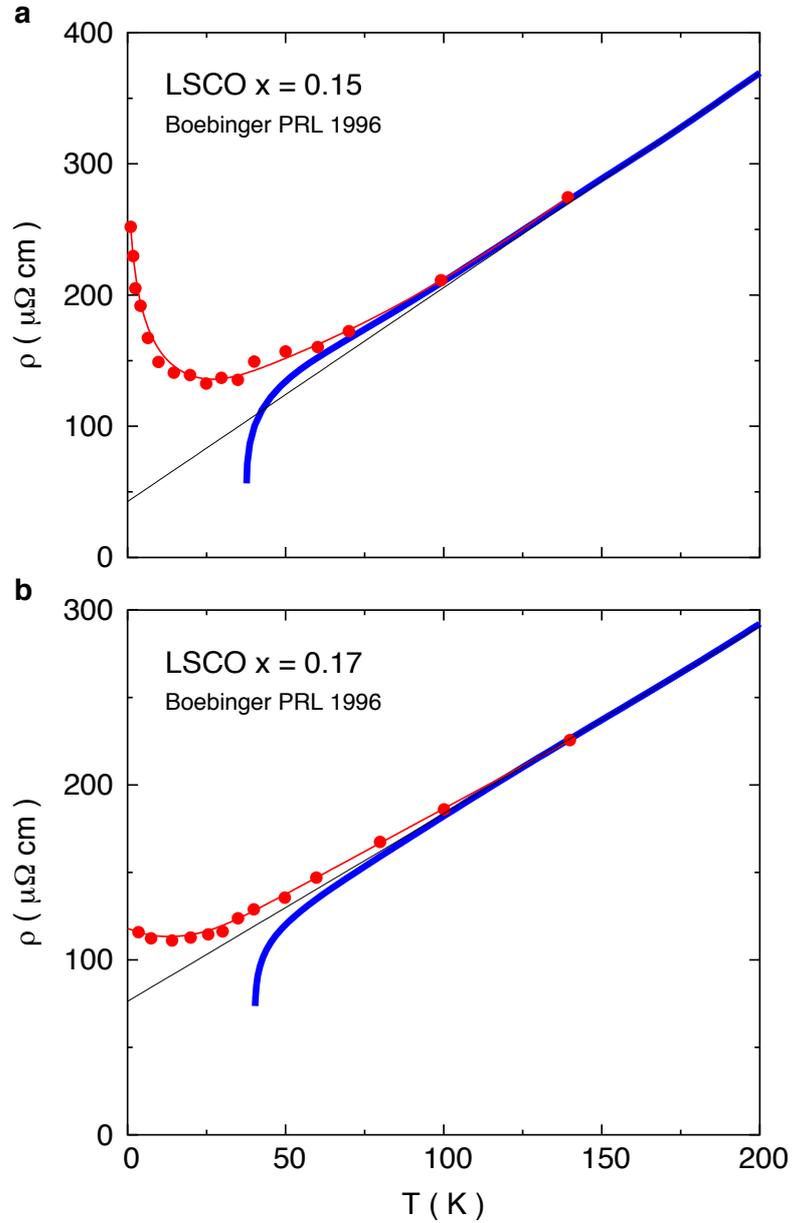

**Fig. S3 | Published resistivity data on LSCO.**

In-plane electrical resistivity ρ of the LSCO samples in ref. 3, with *x* = 0.15 (**a**) and *x* = 0.17 (**b**), as a function of temperature. We assume that *p* = *x*. The blue curve is data in zero magnetic field and the red dots are data in 55 T, both reproduced from ref. 3. The straight black line is a linear fit to the zero-field data at high temperature. Its extrapolation to *T* = 0 gives $\rho_0$ = 43 ± 8 μΩ cm (*x* = 0.15) and $\rho_0$ = 76 ± 10 μΩ cm (*x* = 0.17). The red line is a guide to the eye, whose value at *T* = 0 is ρ(0) = 275 ± 25 μΩ cm (*x* = 0.15) and ρ(0) = 120 ± 10 μΩ cm (*x* = 0.17). The values of ρ(0) and $\rho_0$ are used to calculate $n_\rho$ (see text), plotted as red diamonds in Fig. 4.

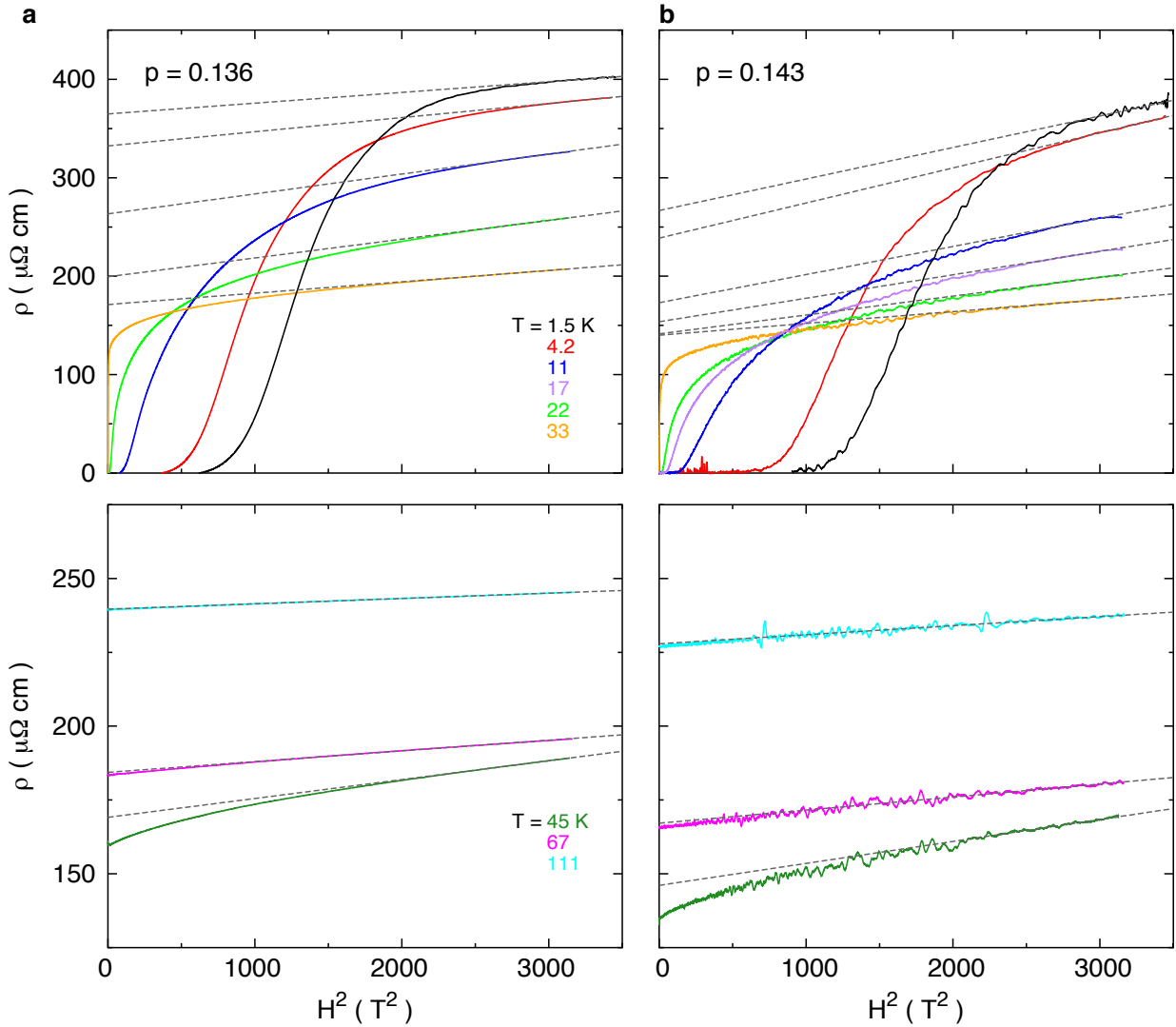

**Fig. S4 | Magneto-resistance vs $H^2$.**

Isotherms of the resistivity ρ of our LSCO samples with $p$ = 0.136 (left panels) and $p$ = 0.143 (right panels), plotted as ρ vs $H^2$. At $T$ = 67 K and 111 K, a linear fit (dashed line) is an excellent fit to the data at all $H$. Applying a linear fit to the other (lower) isotherms at the highest fields yields the dashed lines shown. Extrapolation of these to $H$ = 0 gives values of ρ that are approximately the normal state values at $H$ = 0. These MR-corrected values of ρ are plotted vs $T$ in Fig. S5 (as blue dots).

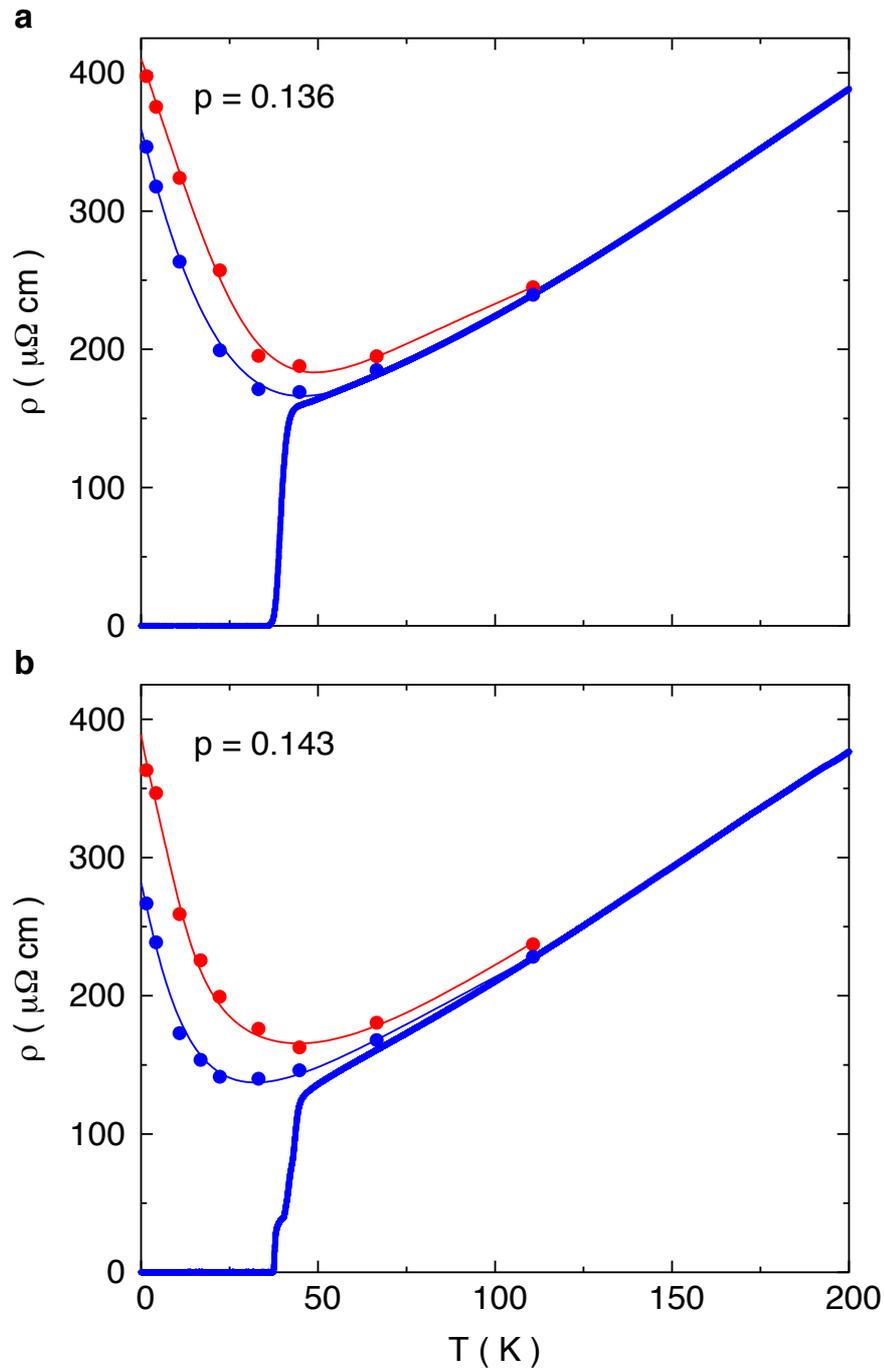

**Fig. S5 | Resistivity corrected for magneto-resistance.**

Resistivity ρ of our LSCO samples with $p$ = 0.136 (**a**) and $p$ = 0.143 (**b**) at $H$ = 55 T (red dots) and corrected for the magneto-resistance (blue dots) by extrapolating the ρ vs $H^2$ data of Fig. S4 to $H$ = 0. The decrease in ρ(0), the value at $T \rightarrow 0$, leads to an increase in the value of $n_\rho$. The raw and corrected values of $n_\rho$ are compared in Fig. S6. The red and blue lines are guides to the eye.

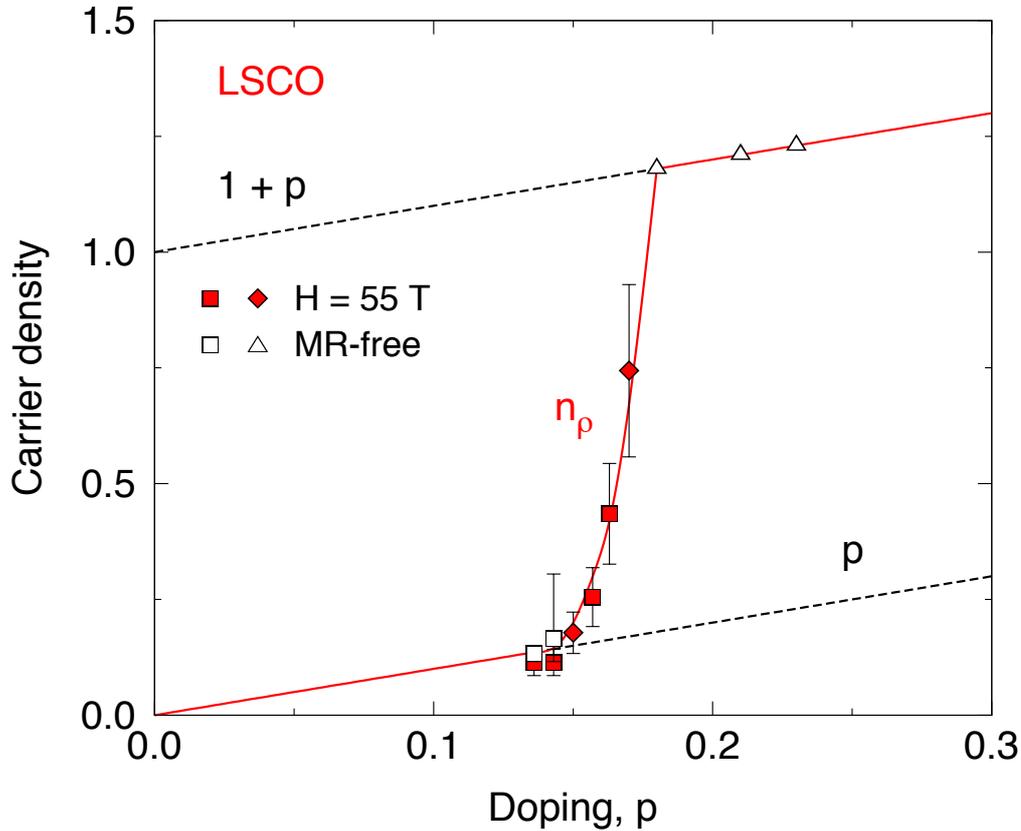

**Fig. S6 | Carrier density with and without magneto-resistance.**

Comparing two ways of estimating the carrier density from $n_p = (1 + p)\, \rho_0\, /\, \rho(0)$: 1) using raw data for $\rho(0)$, measured at $H = 55$ T (full symbols, from Fig. 4a); 2) using a value for $\rho(0)$ that is corrected for the magneto-resistance (open symbols). The two open squares are obtained by extrapolating to $T = 0$ the blue data points in Fig. S5. The error bar on these two data points is defined as follows: the lower bound is the raw value at 55 T; the upper bound is obtained by assuming that the MR in the low-$T$ isotherms of Fig. S2a and S2b is linear ($\rho$ vs $H$), instead of the quadratic dependence ($\rho$ vs $H^2$) assumed in Fig. S4. (The error bar for the data point at $p = 0.136$ is equal to the symbol size.) Even in this extreme case, we see that the MR-free carrier density still drops abruptly below $p^*$, to reach $n_p = p$ at $p \sim 0.14$.

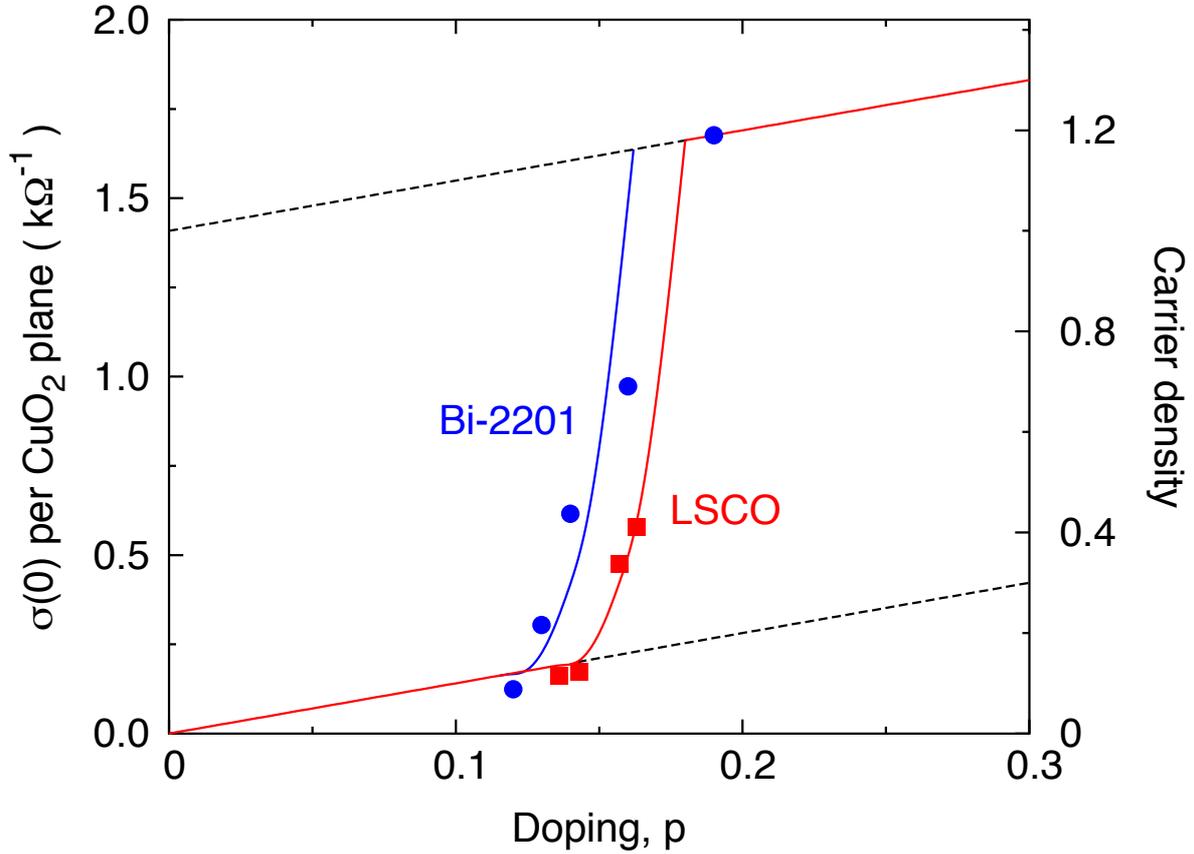

**Fig. S7 | Conductivity and carrier density in Bi-2201 and LSCO.**
Doping dependence of the normal-state conductivity per CuO$_2$ plane in the $T = 0$ limit, σ (0) = $s$ / ρ(0), where $s$ is the inter-plane separation, from high-field measurements of ρ(0) in Bi-2201 at $H = 60$ T (blue dots, left axis; from ref. 4) and in LSCO at $H = 55$ T (red squares, left axis; from Fig. S1). By assuming a doping-independent mobility μ, as observed in LSCO and Nd-LSCO (see text), σ(0) = $n$ $e$ μ becomes a measure of the carrier density $n$. Taking the mobility (at $T = 0$) to be μ = 0.0013 T$^{-1}$, the average value for our LSCO samples, we get the carrier density given on the right axis. The Bi-2201 data point at the highest doping ($p = 0.19$) is such that $n = 1 + p$ (upper dashed line). Below that doping, $n$ in Bi-2201 drops rapidly to reach a value such that $n \sim p$ (lower dashed line) at $p \sim 0.13$. The blue line is parallel to the solid red line that goes through the transition in LSCO (Fig. 4a), starting at $p^* = 0.16$ rather than $p^* = 0.18$. This figure shows that the normal-state resistivity of Bi-2201 (ref. 4) is quantitatively consistent with data in LSCO, both showing a drop in carrier density from $n = 1 + p$ to $n = p$ across $p^*$. Moreover, the width of the transition is roughly the same, within error bars, namely δ$p \sim 0.03$.

# MAGNETO-RESISTANCE AND MOBILITY

In Fig. 4a, we plot the carrier density $n_p$ using the raw data at $H$ = 55 T for our four samples, and compare this to estimates of $n_p$ using published raw data also at 55 T (ref. 3). We see that the agreement between new and old data is excellent.

Now the resistivity of LSCO displays a positive magneto-resistance (MR) in its normal state (Fig. 3), which we should ideally correct for in our estimate of $n_p$. Above $T_c$, the MR goes as $H^2$ (Fig. S4). Assuming that the MR still goes as $H^2$ at lower temperature, we can estimate the normal-state resistivity free of MR by fitting the high-field isotherms to an $H^2$ dependence and back extrapolate to $H$ = 0 (Fig. S4). This gives the blue circles in Fig. S5. Extrapolating these MR-free data to $T$ = 0, we get a value of $\rho(0)$ that is lower than the raw value at 55 T, which in turn yields a slightly larger value of $n_p$. In Fig. S6, we plot the MR-corrected values of $n_p$ (with an upper bound on the correction obtained by assuming that the MR goes as $H$ instead of $H^2$), and see that they are not very different from the raw values. It is clear that correcting for the MR in our data does not alter any of our conclusions.

For $p$ = 0.157 and $p$ = 0.163, it is not possible to correct for the MR as higher fields would be needed to fully reach the normal state at the lowest temperatures (Fig. S2). Nevertheless, we expect the magnitude of the MR to be comparable to the MR at $p$ = 0.136 and $p$ = 0.143 since all samples have a comparable $\rho_0$, and hence a comparable level of disorder scattering, and so a comparable mobility at $T \rightarrow 0$.

Using the magneto-resistance detected in our measurements on LSCO, and the fact that $MR = [\rho(H) - \rho(0) / \rho(0)] \sim (\mu H)^2$, we find that the mobility does not change appreciably across $p^*$. Data in LSCO at $p$ = 0.136 < $p^*$ (Fig. S4) yield $MR$ = 11 % at $T$ = 45 K (near the foot of the upturn) and 23 % at $T$ = 11 K (near the top), for $H$ = 55 T. Data in LSCO at $p$ = 0.23 > $p^*$ (ref. 11), where $\rho(T)$ decreases linearly all the way from 50 K to 1 K, yield a very similar magneto-resistance at $H$ = 55 T, namely $MR$ = 11 % at $T$ = 40 K and 25 % at $T$ = 10 K. This means that the huge upturn in $\rho(T)$ at $p$ = 0.136, which makes $\rho(0)$ roughly 8 times larger than $\rho_0$, involves a negligible change in mobility, and is therefore attributable to an 8-fold drop in carrier density.

**CHARGE DENSITY WAVE PHASE**

We have shown that upon crossing below $p^*$, the carrier density of three different cuprates drops from $n = 1 + p$ to $n = p$ over an interval $\delta p \sim 0.03$ (Fig. 4b). Now, what happens at dopings below $p = p^* - \delta p$ ? At some point CDW order sets in, below a critical doping $p_{CDW}$. This causes a reconstruction of the Fermi surface, whose signature is a drop in the Hall and Seebeck ($S$) coefficients at low temperature, typically such that $R_H$ and $S$ become negative[20,26,27,28]. In the region of CDW order, we cannot necessarily expect to find that $n_p \sim p$, and for obvious reasons we will not find that $n_H = p$.

In YBCO, CDW order is detected by XRD up to $p_{CDW} = 0.16$ (refs. 29, 30), and $R_H$ at low $T$ goes from negative at $p = 0.15$ (ref. 26) to positive at $p = 0.16$ (ref. 6). In Nd-LSCO, CDW order is seen by XRD at $p = 0.15$ (ref. 31), and there is no report of CDW order at higher doping. In Nd-LSCO at $p = 0.15$, $S$ is negative at low $T$ (ref. 32). From the fact that $R_H$ and $S$ are both positive as $T \rightarrow 0$ at $p = 0.20$ (refs. 7, 33), we infer that $p_{CDW} < 0.20$ in Nd-LSCO.

In LSCO, the highest doping at which Seebeck measurements have detected the Fermi-surface reconstruction by CDW order is $p = 0.136$, in our sample with nominal $x = 0.144$ (ref. 20). The fact that the same upturn is observed at $p = 0.136$ and at $p = 0.143$ shows that moving out of the CDW region by increasing $p$ does not remove (or affect) the upturn. Note also that the upturn in $\rho(T)$ at $p = 0.136$ starts at $T^* \sim 150$ K (Fig. S1a), well above the first sign of Fermi-surface reconstruction, seen only below $T \sim 30$ K (ref. 20).

Below $p \sim 0.08$, CDW order disappears in YBCO (ref. 30), LSCO (refs. 19, 20), and Eu-LSCO (ref. 34), a material closely related to Nd-LSCO. At that point, the Fermi surface undergoes another transition, and the Hall and Seebeck coefficients go back to being positive at low $T$ (refs. 20, 27, 28). Interestingly, at $p < 0.08$ one finds that $n_H \sim p$ again[25,35] (Fig. 4a), suggesting that the ground state at $p < 0.08$ may be the same as in the interval between $p_{CDW}$ and $p^*$ (ref. 6).

**LOCALIZATION**

Upturns in the resistivity continue to be observed at $p < p_{CDW}$ (refs. 3, 4), of course, but the quantitative value of the ratio $\rho(0) / \rho_0$ is not expected to be given by $(1 + p) / p$ any more, for two reasons. The first is that the Fermi surface is reconstructed by the CDW order, as mentioned already. The second is that at sufficiently low doping, in particular below the CDW region ($p < 0.08$), the value of $\rho(0)$ in the low-density metal with $n = p$ approaches the condition for localization. Indeed, in a LSCO sample with $\rho_0 = 60$ $\mu\Omega$ cm, we expect that $\rho(0) = \rho_0 (1 + p) / p =$ 800 $\mu\Omega$ cm at $p = 0.08$. This value of $\rho_0$ corresponds to $k_F l = 2$ (ref. 4). Now, if the Fermi surface consists of nodal hole pockets then the total conductivity is two times the conductivity of one hole pocket. The condition for localization, $k_F l = 1$ for each pocket, then becomes $k_F l = 2$ if $k_F l$ is calculated from the measured resistivity. The $\log T$ localization in LSCO (and Bi-2201) is indeed observed only when $\rho(T)$ exceeds a value such that $k_F l = 2$ (refs. 3,4).

Note that localization will also affect $R_H$ at low temperature, so that below $p = 0.08$ the relation $n_H = p$ is observed at temperatures above the localization regime, at $T = 50$-100 K or so[25,35].

## HALL NUMBER IN LSCO

High-field measurements of the Hall effect in thin films of LSCO have found an anomalous behavior of the Hall number[36]: on top of a broad decrease from $n_H \sim 1 + p$ at high $p$ to $n_H \sim p$ at low $p$, consistent with YBCO and Nd-LSCO, a small narrow peak is observed at low temperature immediately below $p^* = 0.18$. We tentatively attribute this peak in $n_H$ to a contribution from electron-like carriers that would partially compensate the drop in the density of hole-like carriers. This would be consistent with scenarios of Fermi-surface transformation that have a narrow intermediate regime of electron and hole pockets immediately below $p^*$, as in the case of antiferromagnetic order[21]. Note that the peak in $n_H$ detected in LSCO is indeed confined to the interval of width $\delta p \sim 0.03$ that is delineated by $n_\rho$ (Fig. 4).